\documentclass[aps,prd,superscriptaddress,twoside,twocolumn,nofootinbib,10pt,%
showpacs,floatfix]{revtex4-1}

\usepackage{amsmath,amssymb}
\usepackage{graphicx,bm}
\usepackage{dcolumn}
\usepackage{epstopdf}
\usepackage{ulem} %% for strike-through
\usepackage[usenames]{color}
\usepackage{subfigure}
\usepackage{float}
\def\be{\begin{eqnarray}}
\def\ee{\end{eqnarray}}

\def\roughly#1{\mathrel{\raise.3ex\hbox{$#1$\kern-.75em%
\lower1ex\hbox{$\sim$}}}}
\def\lsim{\roughly<}
\def\gsim{\roughly>}
\def\la{\langle}
\def\ra{\rangle}
\def\Tr{\rm Tr}
\def\bi{\bibitem}
\def\chiPTs{$\chi$PT$_\sigma$}

\def\chiPT2{$\chi$PT$_2$}
\def\del{\partial}
\def\calL{\cal L}
\def\hatd{\hat{d}}

\allowdisplaybreaks

\begin{document}

\title{Scalar Pseudo-Nambu-Goldstone Boson \\
in  Nuclei and Dense Nuclear Matter}

\author{Hyun Kyu Lee}
\email{hyunkyu@hanyang.ac.kr}
\affiliation{Department of Physics, Hanyang University, Seoul 133-791, Korea}

\author{ Won-Gi Paeng}
\email{wgpaeng@ibs.re.kr}
\affiliation{Rare Isotope Science Project, Institute for Basic Science, Daejeon 305-811,  Korea}

\author{Mannque Rho}
\email{mannque.rho@cea.fr}
\affiliation{Institut de Physique Th\'eorique, CEA Saclay, 91191 Gif-sur-Yvette c\'edex, France}

\date{\today}
%%%%%%%%%%%%%%%%%%%%%%%%%%%%%%%
\begin{abstract}
The notion that the scalar listed as $f_0 (500)$ in the particle data booklet is a pseudo-Nambu-Goldstone (NG) boson of spontaneously broken scale symmetry, explicitly broken by a small departure from an infrared fixed point, is explored in nuclear dynamics. That notion which puts the scalar -- that we shall identify as a ``dilaton" -- on the same footing as the pseudo-scalar pseudo-NG bosons, i.e., octet $\pi$, while providing a simple explanation for the $\Delta I=1/2$ rule for kaon decay,  generalizes the standard chiral perturbation theory (S$\chi$PT) to ``scale chiral perturbation theory," denoted $\chi$PT$_\sigma$,  with {\it one  infrared mass scale for both symmetries}, with the $\sigma$ figuring as a chiral singlet NG mode in non-strange sector. Applied to nuclear dynamics, it is seen to provide possible answers to various hitherto unclarified nuclear phenomena such as the success of {one-boson-exchange potentials (OBEP)}, the large cancellation of strongly attractive scalar potential by strongly repulsive vector potential in relativistic mean field theory of nuclear systems and in-medium QCD sum rules, the interplay of the dilaton and the vector meson $\omega$ in dense skyrmion matter,  the BPS skyrmion structure of nuclei accounting for small binding energies of medium-heavy nuclei, and the suppression of hyperon degrees of freedom in compact-star matter.
\end{abstract}
% 11.30.Qc Spontaneous and radiative symmetry breaking
% 11.30.Rd Chiral symmetries
% 12.39.Dc Skyrmions
% 12.39.Fe Chiral Lagrangians

\maketitle

%

%\tableofcontents

\section{Introduction: The scalar conundrum in nuclear physics}
In  this paper, we explore the applicability to nuclei and nuclear matter the notion recently put forward by Crewther and Tunstall~\cite{CT} that the scalar $f_0(500)$, denoted in what follows as $\sigma$ \footnote{Not to be confused with the fourth component of the chiral four vector $(\vec{\pi},\sigma)$ in 2-flavor linear sigma model.}, is a pseudo-Nambu-Goldstone (pNG for short) boson arising from spontaneous breaking of scale invariance (or more generally conformal invariance) with a {\it small} explicit breaking due to the departure from an IR fixed point {\it and} the current quark mass. Whether an IR fixed point exists in the (matter-free) vacuum of QCD for three flavors ($N_f=3$) is not yet settled.\footnote{For large $N_f$ as in the techni-dilaton approach to Higgs physics, there is an indication, supported by lattice calculations, that such an IR fixed point exists~\cite{yamawaki0}. {The nature of IR fixed point involved is basically different here from the QCD case. Nonetheless} a dilaton chiral perturbation approach similar to what is advocated in \cite{CT} and adopted in medium in this paper has been formulated~\cite{yamawaki}.}   There is, however, neither clear-cut lattice indication nor strong theoretical argument anchored on QCD proper that falsifies it either. In fact, there is a stochastic numerical perturbation calculation that  ``votes for the existence" of an IR fixed point for two massless quark flavors~\cite{horsley}.  Furthermore it is not implausible that such an IR fixed point could be generated in medium as an {\it emergent symmetry} even if it were absent in the matter-free vacuum. In fact there is a compelling reason, as discussed below,  to believe that an IR fixed point of a sort, dubbed ``dilaton-limit fixed point (DLFP),"  can emerge through strong nuclear correlations. The existence of such a symmetry emergence, that we will simply assume in this paper,  has the potential of resolving several long-standing conceptual problems in nuclear physics and offering a possibility to probe strongly compressed neutron-star matter.

A scalar meson with a mass around 600 MeV has figured importantly in nuclear physics since a long time. The only scalar mesonic excitation with such a low mass currently known is $f_0(500)$ with a broad width. Despite the observed large width,  when taken as a local bosonic field, it has fairly successfully accounted for the attractive scalar channel in nucleon-nucleon potentials such as the well-known Bonn boson-exchange potential and, perhaps more significantly, in the highly popular relativistic mean field (RMF) theories for nuclei, nuclear matter and dense (compact-star) matter. There has been a long-lasting controversy as to whether the scalar excitation with such a large width can be considered as a point particle,  depicted in terms of a local field in phenomenological or effective-theory approaches. If it is a hadronic particle, then the question is, what is its QCD structure? Can it be described in quark model, as, say, quarkonium of $(q\bar{q})^m$ complex with $m\geq 1$, a gluonium (glueball) or a mixture thereof? In certain models such as linear sigma model, there is a natural scalar, which is the fourth component of the chiral four-vector in linear sigma model. However it must have a mass $\gsim 1$ GeV to be compatible with the current algebras, so is not relevant far below the chiral scale $4\pi f_\pi\sim 1$ GeV. Furthermore what figures in Walecka model-type~\cite{walecka} relativistic mean field (RMF) theories -- which are highly (perhaps too) successful in heavy nuclei and nuclear matter -- must be a chiral singlet scalar, not the scalar of linear sigma model. If it were the sigma model scalar with a mass $< 1$ GeV as needed in nuclear phenomenology, it would tend to trigger the destabilization of  nuclear matter due to strong attractive many-body forces~\cite{kerman-miller}. As a way out, it has been suggested that in densities near that of nuclear matter, the scalar is a chiral singlet scalar, but at high density as the system approaches chiral restoration, it should transmute to a $\bar{q}q$ configuration that is expected to dominate at the large $N_c$ limit. This feature has been discussed in terms of the DLFP~\cite{DLFP}.

The idea of Crewther and Tunstall that the scalar is a NG boson or more precisely pNG boson of spontaneously broken scale symmetry driven by explicit symmetry breaking provides a simple way to resolve the above ``scalar conundrum" and offers a new perspective on various aspects of nuclear physics. It is the objective of this article to explore the possibilities provided by this point of view. The potential power of the scheme is that the low-lying scalar is a chiral singlet, consistent with the scalar with a low mass that plays a crucial role in RMF theories of nuclei and nuclear matter,  that can be treated on the same footing as the pseudoscalar pseudo-NG bosons -- the pions and kaons -- formulated in what is called  ``scale chiral perturbation theory"~\cite{CT}. Though as yet unproven rigorously, we take this as an {\it assumption that may be valid in medium} and proceed until we are ``hit by a torpedo" if there is any.  We will find along the way an appealing array of observations that go in the right direction, suggesting that we are on the right track.

 In this paper, we formulate a general framework of our approach focusing more on concepts, which will be taken up with practical applications to specific physical systems in a paper that will follow~\cite{pklr}.
\section{The $\sigma$ as a dilaton}
That there is a scalar Nambu-Goldstone boson on par  with a pseudo-scalar Goldsone boson (pion) in low-energy hadronic interactions is an old idea dating way back to late 1960's~\cite{dilaton-old}. What is new here -- and potentially powerful for our purpose if it is proven to be valid -- is that it is formulated in QCD. In terms of the QCD trace anomaly
\be
\theta_\mu^\mu=\frac{\beta(\alpha_s)}{4\alpha_s}G^a_{\mu\nu} G^{a\mu\nu} +(1+\gamma_m (\alpha_s))\sum_{q=u,d,s} m_q\bar{q}q
\ee
there can be exact scale invariance if in the chiral limit $m_q\rightarrow 0$, there is an IR fixed point $\beta(\alpha_{IR})=0$. The suggestion by CT is that such an IR fixed point, nonperturbative in character, is highly plausible and far below from the chiral scale $4\pi f_\pi\sim 1$ GeV,  the QCD $\beta$ function flows along the trajectory leading to the IR fixed point. The scale symmetry associated with the vanishing of $\theta_\mu^\mu$ is then assumed to be spontaneously broken, giving rise to a NG boson. Note that the chiral symmetry is spontaneously broken at the IR fixed point as long as the quark condensate $\la\bar{q}q\ra$ is non-vanishing. That the two spontaneous broken symmetries are intimately locked to each other is the key point of our development. The scalar $f_0(500)$ is identified with that scalar NG boson with the mass generated by spontaneous scale symmetry breaking in the presence of an ``explicit" symmetry breaking encoded in both the departure of $\alpha_s$ from $\alpha_{IR}$ (with a non-zero gluon condensate) {\it and} the current quark mass. Thus the dilaton $\sigma$ joins the pseudo-scalars,  pions and kaons, to form the pseudo-NG multiplet. What ensues is then a more powerful effective field theory that combines both chiral symmetry and scale symmetry with the possibility of doing systematic expansions both in the chiral counting and in the scale counting.

Among the advantages in this approach in particle physics is a simple explanation of the $\Delta I=1/2$ rule for kaon decays that is accomplished by elevating next-to-leading order (loop) terms in three-flavor chiral perturbation theory {$\chi$PT$_3$} into the leading tree order in terms of the $\sigma$ field in $\chi$PT$_\sigma$. What we are interested in is what this scheme with a NG scalar put together with the NG pseudo-scalars does in nuclear phenomena.

{ For applying to nuclear matter, the key degrees of freedom are the nucleon -- proton and neutron -- and the pion, the degrees of freedom figuring in the usual $\chi$PT. Implementing the scalar $\sigma$, on the other hand, requires three flavors including the strangeness. In what follows, however, we will be focusing on non-strange phenomena that take place in nuclear systems, so we will be projecting out the two-flavor sector from $SU(3)$ for most of the consideration, apart from the structure of the $\sigma$.} To apply to strange hadrons, hyperons and kaons need to, and can be, straightforwardly, incorporated.

Before going to the explicit breaking of the scale and chiral symmetries, let us consider the case of exact chiral-scale symmetry. Both the charged axial current $J_\mu^{5a}$ and the dilatation current $D_\mu$ related to the energy-momentum tensor $\theta_{\mu\nu}$ by
\be
D_\mu=x^\nu \theta_{\mu\nu}
\ee will be conserved
\be
\del^\mu J^{5a}_\mu=0, \ \ \del^\mu D_\mu=\theta_\mu^\mu=0.
\ee
The conserved axial current leads to the celebrated Goldberger-Treiman (GT$_\pi$) relation for the axial current
\be
g_A m_N=f_\pi g_{\pi NN}\label{GTpi}
\ee
where $f_\pi$ is the pion decay constant and $g_{\pi NN}$ is the pion-nucleon coupling constant. Similarly, the conserved dilatation current gives the analog that we will call GT$_\sigma$ relation\footnote{It is perhaps worth pointing out that this relation makes a quantitative sense, giving the result in the right ballpark. In relativistic mean-field approaches in nuclear physics, to be commented on in Section \ref{np}, the scalar ($\sigma$) NN coupling required for nuclear phenomenology comes out to be $g_{\sigma NN}\simeq 9-10$ and using $f_\sigma\simeq f_\pi$ suggested on the assumption of an approximate equality in scale between chiral symmetry breaking and conformal symmetry breaking~\cite{CT}, we would get $m_N\simeq 10 f_\pi\simeq 0.84-0.94$  GeV.}.
\be
m_N=f_\sigma g_{\sigma NN}\label{GTs}
\ee
where $f_\sigma$ is the $\sigma$ decay constant and $g_{\sigma NN}$ is the $\sigma$-nucleon coupling constant~\cite{pagels,carruthers}. We note that one can obtain the same relations (\ref{GTpi}) and (\ref{GTs}) by using partially conserved currents, PCAC and PCDC.

One can also write down a scalar analog to Gell-Mann-Oakes-Renner (GMOR) relation (for the pion)
$f_\pi^2m_\pi^2=-\frac 12 (m_u+m_d) \la(\bar{u}u+\bar{d}d)\ra$. It is given by
\be
f_\sigma^2 m_\sigma^2&=& -(4+\beta^\prime)\la\frac{\beta(\alpha_s)}{4\alpha}G^2\ra\nonumber\\
&&- (3-\gamma_m)(1+\gamma_m)\sum_{q=u,d,s} \la m_q\bar{q}q\ra\label{sGMOR}
\ee
where $\beta^\prime=\frac{\del}{\del \alpha_s} \beta(\alpha_s)$. Note that apart from the gluon condensate, this involves the kaon mass in addition to the pion mass that figures in the GMOR relation for the pion.

The GT relations and GMOR relations for $\pi$ and $\sigma$,  being low-energy theorems, are to constitute the leading terms in chiral-scale perturbation theory $\chi$PT$_\sigma$. What makes sense in two-flavor chiral perturbation theory is the small up and down quark masses resulting in small pion mass. However the $\sigma$ mass is comparable to the kaon mass, so in addressing kaon decays, it could make sense to consider them, as is done in the Crewther-Tunstall approach, on the same scale as the pseudo-scalar NG bosons. In addressing nuclear phenomena, on the other hand, one may wonder whether the $\sigma$ mass, $\sim 1/2$ Gev, is small enough to be considered on the same footing as the pion mass. There are, up to date,  no systematic \chiPTs calculations in non-strange sectors, so one cannot say anything, but there are cases where the mass difference of the size in question does not obstruct chiral-perturbative approach. For instance in hidden local symmetry theory, the $\rho$ meson of mass 770 MeV can be treated on the same footing as the pion mass in formulating chiral perturbation theory~\cite{HY:PR}. This aspect is more relevant in nuclear medium since there is a possible symmetry limit, at high temperature or high density, called ``vector manifestation fixed point"~\cite{HY:PR} at which the $\rho$ and $\pi$ become degenerate. As suggested in \cite{DLFP}, the dilaton can also join the pion at what is called ``dilaton limit fixed point."  {\it This points to a possible existence of an IR fixed point of the type suggested by CT in dense medium regardless of whether such a fixed point is present in the QCD vacuum.} These observations suggest that $\pi$, $\rho$, $a_1$ and $\sigma$ could be put on the same footing as discussed in Weinberg's mended symmetries~\cite{weinberg}, where there is a sense in which they all become massless with  $\rho$ and $a_1$ becoming local gauge bosons at the chiral-scale symmetry restoration point.
\section{Scale Symmetric Hidden-Local-Symmetry(HLS) Lagrangian}
We begin by defining the scale dimensions of the fields involved. Under the scale transformation
\be
x\rightarrow \lambda^{-1} x\label{stransform}
\ee
the scale dimension of which is denoted in what follows as $\hatd[x]=-1$, the chiral field $U=e^{2i\pi/f_\pi}$ where $\pi=\frac 12 \vec{\tau}\cdot\vec{\pi}$ transforms\footnote{We will be mainly dealing with non-strange nuclear problems, so we will be mainly focused on  two-flavor systems, but one should keep the strange flavor in mind.}
 \be
 U\rightarrow \lambda^0 U,\label{U}
 \ee
 and the vector field $V_\mu=\frac 12 \vec{\tau}\cdot\vec{V}_\mu$ and the baryon field $\psi$ transform as
 \be
 V_\mu &\rightarrow& \lambda V_\mu,\label{V}\\
  \psi&\rightarrow& \lambda^{3/2}\psi\label{psi}.
 \ee
 Thus their scale dimensions are $\hatd[U]=0$,  $\hatd[V_\mu]=1$ and $\hatd[\psi]=3/2$.
 The derivative and momentum have the dimensions $\hatd[\del]=\hatd[p]=1$. Following \cite{CT}, we write the chiral-scale Lagrangian in terms of scale dimensions in ``chiral-scale expansion" in momentum (equivalently in derivative), quark mass and $\Delta\alpha=\alpha_{IR}-\alpha_s$ with the counting rule
 \be
 {\cal O} (m_q) \sim {\cal O} (p^2)\sim {\cal O} (\del^2)\sim {\cal O}(\Delta\alpha).
 \ee
 This order will be referred to as ``chiral-scale order"  appropriate for $\chi$PT$_\sigma$, to be distinguished from chiral order for $\chi$PT$_2$.
 We write the chiral-scale Lagrangian following \cite{CT},
 as
 \be
 {\cal L}_{\chi PT_\sigma}={\cal L}_{inv}^{d=4}+{\cal L}_{anom}^{d>4}+{\cal L}_{mass}^{d<3}.\label{lag}
 \ee
 The first term has scale dimension 4, the second term is from the trace anomaly and the third term is the quark mass term. The first is scale-invariant and the second and third terms break scale invariance explicitly.  We should note that in the chiral limit in HLS Lagrangian, only covariant-derivative couplings involving pions and vector mesons are allowed. However in the presence of dilaton, non-derivative terms involving the scalar are allowed and hence can produce infrared divergences triggering phase transitions.

\subsection{Meson sector}
Let us first treat the meson sector. Apart from the dilaton $\sigma$, we can limit ourselves to the flavor $SU(2)$ sector as specified above. Baryons will be introduced later.

We start with the HLS Lagrangian with $\rho$ and $\omega$ in flavor $U(2)$ symmetry which, we will assume, is valid in matter-free space as well as in medium\footnote{In the specific applications given \cite{pklr}, there is evidence that the $U(2)$ should break down at high density. For the discussion in this paper, for simplicity, we will however continue with the unbroken symmetry.}. In terms of the chiral counting, the leading chiral order (LCO) is ${\cal O}(p^2)$, consisting of one term quadratic in covariant derivative and a quark-mass term. At the next-to-leading chiral order (NLCO) is ${\cal O}(p^4)$, at which there are 21 normal-parity terms and 3 anomalous-parity terms called ``homogeneous Wess-Zumino" (hWZ) terms. The details\cite{maetal13} are  unilluminating, so will be skipped here. They are also not needed for our discussion. We will drastically simplify the Lagrangian for discussion here.  Fortunately it turns out that for semi-quantitative applications to baryon structure and dense baryonic matter in terms of skyrmions, one can take only one term out of 21 in the ${\cal O}(p^4)$ Lagrangian as explained in \cite{HLMR-multi}. For our purpose, we can simply write
\be
{\calL}_{HLS} (\xi_L,\xi_R,V_\mu)={\calL}_{HLS}^{(2)} + {\calL}_{HLS}^{(4)}+{\calL}_{hWZ}^{(4)}+{\calL}_{mass}\nonumber\\
\label{HLS}
\ee
with $U=\xi^\dagger_L\xi_R$.
The first two terms are of normal parity, the third term is the abnormal-parity hWZ term Lagrangian and  the last term stands for the quark mass term of ${\calL}(p^2)$ in chiral order. The number in the superscript stands for $n$ in ${\cal O}(p^n)$ in the chiral counting. As given, their scale dimensions are\footnote{The vector-meson kinetic energy term is of ${\cal O}(p^2)$ in the HLS/chiral counting~\cite{HY:PR}, so it should belong to the part ${\calL}_{HLS}^{(2)}$, but it is of scale-dimension 4, hence will not figure explicitly in what follows. }
\be
\hat{d}[{\calL}_{HLS}^{(2)}]=2, \ \ \hat{d}[{\calL}_{HLS}^{(4)}]=\hat{d}[{\calL}_{hWZ}^{(4)}]=4\,,
\ee
%{\color{red} where we don't consider the kinetic term of $V^\mu$ included in ${\calL}_{HLS}^{(2)}$ in counting the scaling dimension. }
We will postpone the quark mass term to later.

Now how to put $\sigma$ and arrive at the effective Lagrangian (\ref{lag})?

We follow the standard procedure that exploits the conformal compensator field (or conformalon) $\chi$ related to the dilaton $\sigma$ field by {
\be
\chi=fe^{\sigma/f}\,,
\ee where we introduced the spurion field $f$ of mass dimension 1 and $\hat{d}[f]=0$.
In medium, the ``vacuum" $|0\ra$ changes to $|0_m\ra$, hence the vacuum expectation value changes with density. Therefore at the end of the day, we put $\la\chi\ra= f {\rm e}^{\la \sigma \ra/ f} = f_\sigma$ } for the given vacuum~\footnote{The vacuum change due to medium will be denoted with an asterisk.}.
As a Nambu-Goldstone boson, $\sigma$ transforms under scale transformation by the shift
\be
\sigma\rightarrow \sigma+ f{\rm ln}\lambda
\ee
so
\be
\chi\rightarrow \lambda\chi,
\ee
i.e., $\hat{d}[\chi]=1$.

What we wish to do is to write down an effective $\chi$PT$_\sigma$ Lagrangian in terms of the effective fields $\pi$, $\rho$, ($a_1$ if needed) and $\sigma$ and also $\omega$ that we will put together with $\rho$ in flavor $U(2)$.
%\footnote{The flavor $U(2)$ symmetry for $\rho$ and $\omega$ may not be good at high density although it is fairly good in the matter-free vacuum. For simplicity we will assume it with the caveat in mind.}.
There are two terms coming from the scale dimension 2 term in (\ref{HLS}), one scale invariant and the other scale breaking,
\be
{\calL}^{(2)}_{HLS} (\xi_L,\xi_R,V_\mu,\chi)&=&c^{(2)}_{hls}{\calL}_{HLS}^{(2)} (\frac{\chi}{f_{0\sigma}})^2\nonumber\\
 &+& (1-c^{(2)}_{hls}){\calL}_{HLS}^{(2)} (\frac{\chi}{f_{0\sigma}})^{2+\beta^\prime},
\ee
where $f_{0\sigma}=\la 0|\chi|0\ra$, the vacuum expectation value of the matter-free vacuum. Here and also below, the coefficients $c_{hls}$ are unknown constants.
The first term in the above equation gives rise to the scale invariant first term of Eq.~(\ref{lag}) and the second term to the scale-breaking second term of Eq.~(\ref{lag}). Now one can do the same for the third and fourth terms, each scale-invariant as is. For instance for the third term, one has
\be
{\calL}^{(4)}_{HLS} (\xi_L,\xi_R,V_\mu,\chi)&=&c^{(4)}_{hls}{\calL}_{HLS}^{(4)}\nonumber\\
&+& (1-c^{(4)}_{hls}){\calL}_{HLS}^{(4)}(\frac{\chi}{f_{0\sigma}})^{\beta^\prime}.
\ee
Similar separations can be done in a by-now-obvious way for the $\chi$ kinetic energy term and any other terms that are introduced.

Next to consider is the third term of Eq.~(\ref{lag}), the quark mass term which written in chiral Lagrangian, is of the form\footnote{From here on, we will ignore the anomalous dimension $\gamma_m$.}
\be
{\calL}_{mass}=\frac{f_{0\pi}^2}{4} {\Tr} (MU^\dagger +h.c.)
\ee
where $M$ is the mass matrix with ${\rm diag} M=(m_\pi^2, m_\pi^2, 2m_K^2-m_\pi^2)$ for $SU(3)$ flavor (as needed for the mass formula for the dilaton) and  $f_{0\pi}$ is the pion decay constant in the vacuum. Taking the mass matrix $M$ as a spurion field with scale dimension $\hat{d}[M]=1$, the scale-symmetry-implemented mass term, including the anomalous dimension, is
\be
{\calL}^{d<3}_{mass}=\frac{f_{0\pi}^2}{4} (\frac{\chi}{f_{0\sigma}})^3 {\Tr}(MU^\dagger +h.c.).
\ee
Finally we have non-derivative terms involving the $\chi$ field only
\be
V_\chi=v_1\chi^4 +v_2\chi^{4+\beta^\prime}.
\ee

It is clear that each term comes in with an unknown parameter, so in general there will be much too many unknown parameters to control in the effective Lagrangian.\footnote{For instance, the 21 ${\cal O}(p^4)$ terms will bring in {\it additional} 21 unknown parameters. } Applying to nuclear matter and dense hadronic matter, however, one can rely on the observation that when one sets $\sigma=0$, the resulting theory is the standard chiral perturbation theory ($S\chi$PT) which is accurate in nuclear processes that do not involve scalar degrees of freedom. In terms of the chiral-scale counting given before, the $\beta^\prime$ contribution is of higher order in the bare constants involving the matter fields. Therefore we can safely set $c^{(n)}_{hls}=1$ for $n=2,4$.  This is the widely resorted-to approximation that recognizes that the properties of non-NG fields are affected little by the {\it explicit breaking} of scale invariance, whereas the spontaneous breaking, closely linked to that of chiral symmetry, cannot be ignored.
This is consistent with the standard procedure of introducing dilaton using the conformal compensator~\cite{dilaton-old,grinstein}.

The simplified Lagrangian that we shall consider is
\be
 {\cal L}^{M}_{\chi PT_\sigma} (\pi,\chi,V_\mu) & \approx&
  {\calL}_{HLS}^{(2)} (\frac{\chi}{f_{0\sigma}})^2 + {\calL}^{(4)}_{HLS}\nonumber\\
   &+& c_{hWZ} {\calL}^{(4)}_{hWZ} + (1-c_{hWZ}) {\calL}^{(4)}_{hWZ}(\frac{\chi}{f_{0\sigma}})^{\beta^\prime}\nonumber\\
   & +&
  \frac{f_{0\pi}^2}{4} (\frac{\chi}{f_{0\sigma}})^3 {\Tr}(MU^\dagger +h.c.)\nonumber\\
    &+&\frac 12 \del_\mu\chi\del^\mu\chi + v_1 {\chi}^4+v_2 {\chi}^{4+\beta^\prime}.\label{cslagM}
 \ee
{ Note that we have retained in this Lagrangian the potential effect of the explicit scale symmetry breaking in the hWZ term that brings in the $\omega$ degree of freedom. This is because the presence of $\omega$ in dense skyrmion matter makes a drastic effect on the behavior of hadrons in dense matter which requires a basic change in the structure of the anomalous-parity term in the presence of density.}

 The possible role of the explicit scale  symmetry breaking in the potential will be discussed below. The potential is given by the non-derivative terms
%\footnote{The presence of the baryon field contributes $\frac{\chi}{f_{0\sigma}}m_B\bar{\psi}\psi$ %to the potential, which should be added. See Eq.~(\ref{cslagB}).}
 \be
- V=v_1{\chi}^4+v_2{\chi}^{4+\beta^\prime} +\frac{f_{0\pi}^2}{4} (\frac{\chi}{f_{0\sigma}})^3 {\Tr}(MU^\dagger +h.c.).\nonumber\\
\label{CTpotential}
\ee
 This gives the trace of the energy-momentum tensor{
 \be
  \theta_\mu^\mu=&& - \beta^\prime v_2 {\chi}^{4+\beta^\prime} \nonumber\\
   && + \frac{f_{0\pi}^2}{4}{\Tr} (MU^\dagger +h.c.)(\frac{\chi}{f_{0\sigma}})^3.
 \ee}
The potential is minimized  by
 \be
&&[ 4v_1{\chi}
+v_2(4+\beta^\prime){\chi}^{1+\beta^\prime}\nonumber\\
%+ m_V^2 \frac{\chi^{-1}}{f_{0\sigma}^2} V_{\mu}^{a}V^{a\,\mu} - m_\psi \frac{\chi^{-2}}{f_{0\sigma}}\bar{\psi}\psi \nonumber \\
&&+3\frac{f_{0\pi}^2}{4f_{0\sigma}^3}{\Tr}(MU^\dagger +h.c.)]_{\chi=f_{\sigma}} =0\, .\nonumber\\
\ee

This determines $f_{\sigma}=\la\chi\ra$ in the {\it bare Lagrangian} as a function of $\beta^\prime$, $v_1$, $v_2$ and the quark mass matrix $M$, which carry  information on the intrinsic density dependence through the condensates.  At present, only the mass matrix $M$ { in matter free space} out of four constants is known, so we are not able to make any estimate.  We will suggest as in \cite{CT} that $f_\sigma\approx f_\pi$ in the matter-free vacuum and also in medium.

There is an important point to note here. Consider the chiral limit. The potential is then
\be
V=v_1{\chi}^4 +v_2 {\chi}^{4+\beta^\prime}.
\ee
If there were no explicit scale symmetry breaking, i.e., $\beta^\prime=0$, then the potential would be of the form $\kappa\chi^4$. In this case, Poincar\'e-4 invariance requires that $\kappa$ be zero~\cite{fubini,serra}, in which case $\la\chi\ra=f_\sigma$ is undetermined because the potential is flat. This is because  spontaneous scale symmetry breaking is possible {\it only} in the presence of an explicit breaking, i.e., either $\beta^\prime\neq 0$ or $M\neq 0$ or both are nonzero,  a feature that differentiates scale symmetry from global symmetries~\cite{freund-nambu}.

\subsection{Baryon Sector}
One natural way justifiable in the large $N_c$ limit of bringing in baryons into the chiral-scale Lagrangian (\ref{lag}) or (\ref{cslagM}) is, as discussed in \cite{maetal13}, to generate them as solitons, i.e., skyrmions, in the mesonic Lagrangian.  We will discuss what one can expect to learn in this approach in a later section. Here we simply introduce baryons as massive matter fields coupled to $\pi$ and $V_\mu$ in a hidden local symmetric way. As noted, the baryon field has $\hat{d}[\psi]=3/2$ and the covariant derivative has $\hat{d}[D_\mu]=1$, so the leading chiral order ${\cal O}(p)$ term involving bilinear fermion fields and $(\pi,V_\mu)$ fields gives the Lagrangian a scale-dimension 4. Thus the baryon couplings to $\pi$ and $V_\mu$ are scale invariant. The baryon mass term, however, has $\hat{d}[m_B\bar{\psi}\psi]=3$ so it needs to be multiplied by $\frac{\chi}{f_{0\sigma}}$ to make it scale-invariant. As in the meson sector, one could also account for the effect of quark-gluon mixing in the baryon sector, incorporating the effect of scale-symmetry explicit breaking $\beta^\prime$. It is clear from the scalar GT mass formula for nucleon (\ref{GTs})  that  the explicit symmetry breaking can be ignorable and the nucleon mass is dominated by gluon effects~\cite{CT}. There is an evidence for this in the study of dense baryonic matter where the nucleon mass remains non-vanishing and large when $\la\bar{q}q\ra$ goes to zero~\cite{maetal13}.  We have then
\be
{\cal L}^{B}_{\chi PT_\sigma} (\psi,\pi,\chi,V_\mu)\approx {\cal L}_{HLS,baryon} - \frac{\chi}{f_{0\sigma}} m_B \bar{\psi}\psi\, .\label{cslagB}
\ee
%As noted above, the last term should be added to the potential (\ref{CTpotential}), { which will %give additional density dependence to $\la\chi\ra$ on top of  the intrinsic density dependence  in %$M, \beta', v_1, v_2$. }
\section{Applications in nuclear physics}\label{np}
\subsection{Nuclear $\chi$PT$_\sigma$}
  Up to date, the effective field theory approach to nuclear dynamics anchored on chiral symmetry has been chiral perturbation theory using only the pion field, either with or without baryons fields, the latter via solitons. This is the standard (two-flavor) chiral perturbation theory S$\chi$PT (and baryon $\chi$PT (B$\chi$PT) when baryons are included).  The S$\chi$PT is based on the assumption that more massive meson degrees of freedom than pions can be integrated out with their effects inherited in the parameters of S$\chi$PT Lagrangian. This is evidently justifiable for the vector mesons $\rho$ and $\omega$. It has been unclear, however, where the scalar degree of freedom that has figured in phenomenological nuclear potential models such as Bonn potential and also in a class of density-functional models, in particular, relativistic mean field (RMF) approach initiated by Walecka, fit into the chiral perturbative scheme. An implicit assumption  was that a scalar of mass around 600 MeV can be treated as a {\it local scalar field}. It has been difficult to find a justification for this since the only low-lying scalar presently known in the particle data booklet is $f_0(500)$ but it has a large width, of the order of the mass. In S$\chi$PT, such scalar excitations are to be generated at higher order chiral perturbation, invoking dispersion relations, higher-loop unitarity corrections etc. That such S$\chi$PT with baryons included, i.e., B$\chi$PT, seems, at higher chiral order, to successfully describe nuclear matter~\cite{HRW-GEB} indicates that multi-loop corrections do generate appropriate scalar attraction necessary for binding in nuclei and nuclear matter.

   The power of the CT approach is that the pseudo-dilaton is to capture the dynamics of the scalar channel at the leading (tree) order together with the pion, relegating loop terms to small corrections. Having { $\sigma$} and $\pi$ (and also $K$) on the same footing at the same chiral-scale order would make the calculation of nuclear potentials vastly simpler and more efficient. The role of the dilaton would resemble the scalar exchange in the popular Bonn one-boson-exchange two-nucleon potential. It has the advantage of enabling one to make systematic error estimates of higher order terms, a power that is lacking in phenomenological potential models. It will also affect medium-range three-body potentials mediated by $\sigma$ exchanges. Whether or not this scheme supersedes the S$\chi$PT in nuclear process needs to be investigated, clearly a brand-new open field in nuclear physics. In what follows, we will describe certain features that show promise for novel developments.
 \subsection{Relativistic mean field treatment of hidden local symmetry with scale invariance}\label{RMF}
 In this subsection we treat the chiral-scale Lagrangian with the baryon field explicitly  incorporated. In the next subsection we will introduce baryons as skyrmions from the chiral-scale Lagrangian.

 The Lagrangian we take is given by the sum of (\ref{cslagM}) and (\ref{cslagB}),
 \be
 {\calL}_1={\cal L}^{B}_{\chi PT_\sigma} (\psi,\pi,\chi,V_\mu) + {\cal L}^{M}_{\chi PT_\sigma} (\pi,\chi,V_\mu)\,.\label{L1}
 \ee
 The dilaton potential defined by non-derivative terms  Eq.(\ref{CTpotential}) should be  changed in the presence of baryons to
 \be
- V &=& v_1{\chi}^4+v_2{\chi}^{4+\beta^\prime} +\frac{f_{0\pi}^2}{4} (\frac{\chi}{f_{0\sigma}})^3 {\Tr}(MU^\dagger +h.c.) \nonumber\\
& & - \frac{\chi}{f_{0\sigma}} m_B \bar{\psi}\psi\
\label{CTpotentialB}
\ee
As mentioned, the dilaton potential, Eq.(\ref{CTpotentialB}), is to be minimized at the medium-free vacuum ($\la\bar{\psi}\psi\ra=0$) at which $\la\chi\ra_{n=0}=f_{0\sigma}$. Let us consider the vacuum is changed to one in which baryon density $n >0$,  call it ``m(edium)-vacuum" and denote it $|0_m\ra$. Then the dilaton condensate in that m-vacuum must be changed to $\la 0_m |\chi| 0_m \ra= \la\chi\ra^*\equiv f_\sigma^*$ as a solution of Eq.~(\ref{CTpotentialB}). In this m-vacuum, shift the dilaton field
 \be
 \chi=f_\sigma^*+\chi^\prime
 \ee
and substitute this into (\ref{L1}). We get two terms, one without $\chi^\prime$ and one with. The first one gives back the original baryon HLS Lagrangian (without conformal compensator field) with two parameters changed as
\be
f_\pi&\rightarrow& f_\pi^*=\frac{f_\sigma^*}{f_{0\sigma}}f_{0\pi}\label{fpi}\\
m_\pi &\rightarrow& m_\pi^*= \sqrt{\frac{f_\sigma^*}{f_{0\sigma}}}m_\pi\label{mpi}
\ee
with the hidden gauge coupling $g$ and $a$ remaining unchanged\footnote{$a$ is a parameter in HLS, $a=f_{\bar{\sigma}}/f_\pi$, where $\bar{\sigma}$ is the would-be Nambu-Goldstone boson that's eaten up by the $\rho$ to become massive via Higgs mechanism. }
%{\color{blue} It should be noted that there is also the intrinsic density dependence of $m_{\pi}$ from the QCD matching discussed in section V. }}.
As a consequence the vector meson and nucleon masses will be modified as\footnote{These are the scaling obtained in 1991~\cite{br91}.}
\be
m_V=\sqrt{a}f_\pi g\rightarrow m_V^*=af_\pi^*g=\frac{f_\sigma^*}{f_{0\sigma}}m_V\label{mv}\\
m_N\rightarrow m_N^*=\frac{f_\sigma^*}{f_{0\sigma}} m_N.\label{mn}
\ee

{ We should mention briefly how the density-dependent $f_\sigma^*$ -- which figures crucially in low-density regime -- is determined. More precise discussions will be given below in Section \ref{matching}.

In principle if we know how $f_\sigma^*$ depends on QCD variables, then its ``intrinsic density dependence (IDD)" will be known in terms of the density dependence of those QCD variables that enter. It will be encoded in the effective in-medium dilaton potential, which is presently not known without precise knowledge on the scale symmetry breaking, both explicit and spontaneous. For low density, one can, however, exploit the relations (\ref{fpi}) and (\ref{mpi}) linking the $\sigma$ decay constant to the pion decay constant, since the pion properties can be extracted from deeply bound pionic systems that can be accessed experimentally.}

Now the Lagrangian with $\chi^\prime$ describes $\sigma$ coupling, in a way consistent with scale symmetry, to the vector mesons and the nucleon, with the modified parameters given above.  Particularly important to nuclear dynamics as explained below is the $\sigma NN$ coupling\footnote{For the chiral-scale Lagrangian to reproduce, at  tree order, the low-energy theorem (\ref{GTs}), the sign for the coupling constant is chosen so that $g_{\sigma NN} >0$.} {
\be
{\calL}_{\sigma NN}=-\frac{m_N}{f_{0\sigma}} \bar{\psi}\psi\chi^\prime =g_{\sigma NN}\bar{\psi}\psi\sigma+\cdots
\ee }
where the ellipsis stands for higher fluctuating $\sigma$ fields and the GT$_\sigma$ relation (\ref{GTs}) is used in the last equality. It is worth noting that the $\sigma NN$ coupling remains unscaling at the order considered.

The resulting Lagrangian is then applicable to the description of nuclear dynamics in a baryonic medium with density $n$. It is easy to see that when expanded to lowest order in the fields with only the baryon $N$, scalar $\sigma$ and isoscalar vector field $\omega_\mu$ appropriate for symmetric nuclear matter\footnote{Both $\rho$ and $\pi$ fields can be brought in HLS symmetric way for asymmetric nuclear systems.}, it reproduces the Walecka's model Lagrangian~\cite{walecka} with, however, the mass parameters given a prescribed density dependence,{
\be
{\calL} &=& \overline{N}[i\gamma_\mu(\del^\mu+ig_v\omega^\mu)-m_N^*+g_{\sigma NN}\sigma]N\nonumber\\
 &-& \frac14 F_{\mu\nu}^2+\frac{{m^*_\omega}^2}{2}\omega^2 + \frac 12(\del_\mu\sigma)^2 -\frac{{m_\sigma^*}^2}{2}\sigma^2. \label{waleckap}
 \ee}
 The in-medium $\sigma$ mass $m_\sigma^*$ is subtler and requires a lot more care. We will return to it in Section \ref{sigmamass}.  Apart from $m^\star_\sigma$, this Lagrangian has {implicit} density dependence only in one quantity, $\Phi=f_\sigma^*/f_{0\sigma}$, reflecting the interlocking of scale and chiral symmetries. As mentioned in Section \ref{matching}, the matching of the effective Lagrangian to QCD at the matching scale endows an ``intrinsic density dependence" inherited from QCD in the ``bare" parameters of the Lagrangian which manifests itself at a density above that of nuclear matter $n_0$. For the moment, we ignore this at least up to $n_0$. This can be justified as discussed in \cite{pklr}.

As mentioned,  without knowing the precise mechanism for the explicit breaking of scale symmetry, we cannot calculate the in-medium properties of the parameters $v_{1,2}$, $\beta^\prime$ etc. of the potential (\ref{CTpotential}). However $f_\pi^*/f_{0\pi}$ can be measured in certain experiments such as deeply bound pionic systems~\cite{yamazaki}. If we ignore for the moment the IDD in other parameters in (\ref{waleckap}) which will be justified later, (\ref{fpi}) allows one to apply the mean field to the Lagrangian (\ref{waleckap}) and see how it fares for nuclear matter. It in fact is found to work well. This has been done by adjusting the ``bare" (that is, un-starred) parameters of the Lagrangian in the standard manner to fit nuclear matter properties~\cite{song}. This model improves, more significantly, on the original linear model of Walecka in that the density dependence of the parameters removes one of the defects of Walecka's linear model, namely the compression modulus which is too big, roughly by a factor of 4, compared with the empirical value. In relativistic mean field models, the remedy is made by incorporating higher field terms in consistency with the strategy, e.g., naturalness etc,  of effective field theory and their condensates ``renormalizing" the parameters of the Lagrangian. The Lagrangian ({\ref{waleckap}) is found to nearly fully implement this nonlinear effect by the dilaton condensate reflecting the role of the sliding m-vacuum.

\subsection{Vector and scalar mean field}
The relativistic mean field (RMF) approaches, belonging to the general paradigm of density functional theory and generalizing Walecka linear mean-field model, are found to work remarkably well for finite nuclei and nuclear matter~\cite{RMF}, and are now being applied, with some success,  to highly dense matter expected to be found in the interior of compact stars. The most outstanding feature of RMF approaches as applied to nuclear matter is the large cancellation of the scalar mean field energy-- which is attractive -- and the vector mean field energy -- which is repulsive~\cite{walecka}. The binding energy that results, $\sim 16$ Mev, is tiny compared with the respective mean field energies $\sim 300$ MeV. We can capture what's happening here in terms of the nucleon self-energies in nuclear matter obtained in in-medium QCD sum rule calculations. The detailed analysis made in \cite{cohenetal} can be succinctly summarized by the nucleon self-energies in medium at density $n$
\be
\Sigma_s/m_N&\approx& - \frac{\sigma_{\pi N} }{m_\pi^2 f_\pi^2} n\, ,\\
\Sigma_v/m_N&\approx& \frac{8m_q}{m_\pi^2 f_\pi^2}n
\ee
where $\sigma_{\pi N}$ is the $\pi N$ sigma term and $m_q$ is the average light-quark mass. It is found numerically that at nuclear matter density,  $\Sigma_s/m_N\approx - (0.3-0.4)$ and $\Sigma_v/m_N\approx (0.3-0.4)$. In terms of the $\sigma$-implemented hidden local symmetry Lagrangian, these self-energies correspond to two tadpole diagrams (Fig. 1), one involving a $\sigma$ pole giving the scalar attraction and the other an $\omega$ exchange leading to the vector repulsion. Both are big but the sum is tiny. We will see below in the description relying on skyrmion structure that this feature emerges in what appears to be a totally different mechanism.

\begin{figure}[!t]
\includegraphics[width=0.3\textwidth]{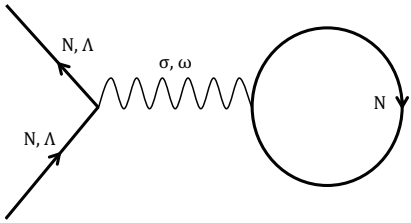}
\caption{Tadpole diagram for self-energies for the nucleon $N$ and the hyperon $\Lambda$ in medium. The loop corresponds to the nucleon scalar density $n_s=\la\bar{N}N\ra$ for coupling to $\sigma$ and the nucleon number density $n=\la N^\dagger N\ra$ for coupling to $\omega_0$.}
\label{fig1}
\end{figure}

The near cancellation of the QCD sum-rule self-energies supports the near cancellation found in RMF analyses. This cancellation will persist at higher density above the normal, however becoming more intricate  due to the scaling properties of the dilaton and the $\omega$ meson at higher densities. This feature must play a crucial role in generating repulsion in the EoS needed to lead to the massive neutron stars observed recently. An explicit calculation in \cite{pklr} does indeed confirm this prediction.

The simplest, though far from rigorous, way to understand the impressive success of the RMF approach in nuclei and nuclear matter is that the system is at the Landau Fermi liquid fixed point~\cite{shankar} and that the RMF approach captures the Landau Fermi-liquid fixed point theory. This suggests that the RMF approximation could be valid in dense medium as long as the system is in Fermi liquid. We will comment in the Discussion section on the possibility that the Fermi-liquid structure could break down at high density as it happens in certain condensed matter systems. This could have drastic consequences on the EoS of compact-star matter, an aspect that has not yet been explored.
\section{Matching to QCD, vector manifestation and mended symmetries}\label{matching}

%{\color{blue} HKL: In this introductory section,  two aspects  of QCD matching are discussed. (1) Far away from VM fixed point, intrinsic density dependence is negligible such that    $g$ and  $a$ are almost constant.  Then it is likely  the intrinsic density dependence of  $v_1, v_2$ and $M$ is almost negligible if matching would have been done   with ${\cal L}^{M}_{\chi PT_\sigma} (\pi,\chi,V_\mu)$. Am I right? Density independent meson mass parameter $M$ seems to be OK.  However  at least one of $v_1$ and  $v_2$    should depend on the density to give the proper density dependent vev of $\chi$ as a solution of Eq.(27) or $\Phi$ factor.  For the moment I could not find a good resolution of this problem?   (2) Near the VM point,  the intrinsic dependence is argued, from QCD matching, to be dramatically different from that of far-away-from VM point. If we accept this, then the method presented above with Eq. (\ref{L1}) should be  ultimately be modified at some density not too high above $n_0$. (3) Using the pNG nature,  the mass of  $\sigma$ at lower density is found to stay almost constant, like $g$ and $a$,  and  in chiral-scale limit(assuming at the same  VM density)  possibly at higher density it becomes massless degenerate with  $\pi$, $\sigma$, $\rho$ and $a_1$ for Weinberg's mended symmetry. (4) Throughout here the effect of  RG-decimation has not been presented  even qualitatively although it is necessary to obtain low energy values of parameters. Is it better to  add some more discussion or not necessary?}

While the method presented above with (\ref{L1}) that can be given some theoretical and phenomenological support near nuclear matter density could be extended to somewhat above the equilibrium density, there are reasons to expect that it will ultimately fail within the hadronic phase at some density not too high above $n_0$. One reason is that the Fermi-liquid structure could be broken down at high density. This will be discussed later. Another reason has to do with the ``intrinsic density dependence" (IDD) that the effective field theory inherits from QCD at the scale where effective field theory is matched to QCD. This effect is embedded in the parameters of the Lagrangian at the scale the effective field theory is defined. While the density dependence brought in by the $\Phi$ factor is more less under control, most likely up to nuclear matter density and perhaps slightly above, ignoring the IDD in other parameters of the effective Lagrangian cannot be valid at higher density, particularly as the chiral restoration density is approached. Among others, the most compelling reason is the vector manifestation property of HLS~\cite{HY:PR}.

{As alluded to above without precision, the ultraviolet completion of the effective Lagrangian, a requirement for an effective field theory for QCD, is made by matching the correlators of the EFT -- the dialton-implemented HLS ($s$HLS for short) -- to those of QCD at a matching scale $\Lambda_M$ slightly below the chiral scale $\Lambda_\chi\sim 4\pi f_\pi$. From it, the parameters of the ``bare" $s$HLS Lagrangian inherit from QCD the dependence on non-peturbative quantities, i.e., the condensates, $\la\bar{q}q\ra$, $\la G^2\ra$, etc. defined at the scale $\Lambda_M$ in the matter-free vacuum. In applying this Lagrangian to in-medium systems at low energy, two operations are required. One is to take into account the vacuum structure caused by density and the other is to RG-decimate from $\Lambda_M$ to the low-energy scale where physics is done. The first is captured by the density dependence of the condensates that we call ``intrinsic density dependence (IDD)," mainly in the quark condensate since the gluon condensate is insensitive to density~\cite{cohenetal}\footnote{The decrease of the gluon condensate is estimated to be $\sim 5\%$ at nuclear matter density. Given the large uncertainty, the effect of this change could be ignored in our analysis.}, and the second is to do quantum (loop) corrections using the density-dependent parameters.

One can classify two classes of intrinsic density dependence (IDD). One involves the matter fields, vector mesons and baryons. It comes from matching the isovector vector and axial vector correlators at the scale $\Lambda_M$ (at $n=0$). We shall call this IDD$_m$. The other involves pNG bosons, $\sigma$ and $\pi$. It is gotten from matching the trace of energy-momentum tensor. We shall call the resulting density dependence IDD$_{pNG}$. The IDD$_{pNG}$ embodies the common IR scale for the scale and chiral symmetries leading to the relations (\ref{fpi}) and (\ref{mpi}). Due to the interplay between the two symmetries, IDD$_{pNG}$ affects also the matter fields giving rise to the scalings (\ref{mv}) and (\ref{mn}).  The masses and coupling constants involving the matter fields carry, in addition,   the IDD$_{m}$.

In doing the mean field approximation with, e.g., (\ref{waleckap}), the IDD$_{pNG}$  effect  is included via the dilaton condensate, but the IDD$_m$ effect, which is difficult to calculate  except near the VM fixed point,  is not taken into account. Fortunately there is a good reason to believe that the latter effect could be  negligible far away from the chiral restoration point. Thus for instance, the $\rho$ meson mass is to scale with $f_\sigma^*$, but not with the gauge coupling $g$ nor with the parameter $a$.  In fact,  it has been verified phenomenologically that the density dependence is fairly well captured by the dilaton condensate, at least, up to nuclear matter density~\cite{BR:DD}. But this cannot continue to be valid as the matter approaches the chiral restoration density $n_c$. Among others, the most compelling indication for the breakdown comes from the vector manifestation property of HLS ~\cite{HY:PR}.}

It is shown in \cite{HY:PR} that as the chiral symmetry gets unbroken, by temperature or density or other mechanisms so that the quark condensate goes to zero, the $\rho$ mass should approach zero (in the chiral limit) as
\be
m^*_\rho\sim g^*\sim\la\bar{q}q\ra^*\rightarrow 0.\label{HLSg}
\ee
Thus while the $\rho$ mass drops at low density mainly due to the decrease of $f^*_\pi$ with the gauge coupling $g$ and $a$ remaining constant,  at high density it is the dropping gauge coupling that should take over as indicated in (\ref{HLSg}). It is found also that the pion decay constant does remain more or less constant starting from a density $\sim (2-3)n_0$ up to near the chiral transition. For this, the explanation could be that the dilaton decay constant is controlled at high density by the gluon condensate with the quark condensate strongly suppressed and the locking of the scale and chiral symmetries makes the pion decay constant behave the same as the dilaton decay constant. There is an effective change in the role for precursor for chiral symmetry restoration between $f^*_\sigma$ (or equivalently $f^*_\pi$) and the hidden gauge coupling $g^*$ at $n\sim 2n_0$. This suggests that the $\rho$ mass, not $f_\pi^*$,  plays the role for order parameter for chiral restoration. A similar phenomenon takes place also for the nucleon mass. We will comment on this in Discussions.
\subsection{The mass of  pseudo-Nambu-Goldstone bosons}\label{sigmamass}
Now what about the dilaton mass in dense medium?
This is an extremely intricate matter, a clear answer to which requires a deep understanding of both explicit and spontaneous symmetry breaking of scale invariance, which is at present missing. The only thing one can say is that since the dilaton is a pseudo-NG boson, its behavior could be closer to that of the pion than to the vector meson and the nucleon.

To get some insight, let us proceed with the property of pion in medium, fairly well studied both in experiments and in theory. For the latter, standard chiral perturbation theory is surely  suitable. Let us see what we can expect.

From Eq.~(\ref{mpi}) we have
\be
\frac{m_\pi*}{m_\pi}\approx \sqrt{\frac{f_\pi^*}{f_\pi}}\, .
\ee
 It has been established from the study of deeply bound pionic systems, that $(f_\pi^*/f_\pi)^2\approx 0.64$ at nuclear matter density $n=n_0$~\cite{yamazaki}. This gives $\frac{m_\pi*}{m_\pi}\approx 0.9$. However this is not the whole story in nuclear matter because what is observed in nature \cite{yamazaki} contains also  pion fluctuation effects. The latter can be computed reasonably well with the chiral Lagrangian containing ``bare" parameters. It comes out to give $\sim 10\%$ increase to the mass at $n=n_0$~\cite{weise}. To compare with the experiment, one has to combine the two. The net effect is $0.9\times 1.1\approx 1$. This means that the pion mass should remain unshifted by matter, at least up to nuclear matter density. This seems to be somewhat at odds with the measurement of the pion mass shift which sees a small increase in mass but this could be due to uncertainty in the interpretation. It could depend on how the in-medium mass for the pion is defined -- which is not unique -- in the analysis of the experimental data.

It is not feasible at the moment to exploit a similar strategy for the in-medium mass of a dilaton. This is because the mechanism for the explicit scale symmetry cannot be pinned down given the numerous unknown constants such as $v_1$,$v_2$ and $\beta^\prime$ etc. in the effective Lagrangian.  {We can, however, make an indirect argument based on QCD sum-rule analyses of gluon condensates using Eq.~(\ref{sGMOR}). Ignoring the pion mass relative to the kaon mass, we have in medium \footnote{ Given that the process must take place near the IR fixed point, one can expand $\beta$ near the fixed point and set $\beta(\alpha_s)\approx (\alpha_{IR}-\alpha_s)\beta^\prime(\alpha_{IR})$. Note it is the departure from the IR fixed point that represents the explicit breaking of scale symmetry.}{
\be
{m^*_\sigma}^2 \approx {f^*_\sigma}^{-2} \left[(-(4+\beta^\prime) \la\frac{\beta(\alpha_s)}{4\alpha_s}G^2\ra^* + 3 {m_K^*}^2 {f_\pi^{\ast}}^2)\right].
\ee}}
 From QCD sum rules, one knows that the gluon condensate is little affected by density~\cite{cohenetal}.  Since { $\la\frac{\beta(\alpha_s)}{4\alpha_s}G^2\ra^*\sim -\beta^\prime v_2\la\chi^{4+\beta'}\ra^*$}, this indicates that $f^*_\sigma=\la\chi\ra^*$ is also insensitive to density, particularly above some density\footnote{It is not yet clear how to pin down this density in this formalism. In the skyrmion crystal description of dense matter~\cite{half-skyrmion}, this behavior, reflected in the pion decay constant $f_\pi^*\approx f_\sigma^*$, appears at the  density at which skyrmions fractionize into half-skrymions,  $n_{1/2}
  \sim 2n_0$. We equate the two densities in \cite{pklr} in applying the formalism to the EoS of dense matter.}.  Also the s-quark condensate $\la\bar{s}s\ra$  is expected to be  less sensitive to density than the light-quark condensate, so the kaon mass would also be little affected for not too high density. Thus the in-medium $\sigma$ mass is affected primarily by the $\sigma$-decay constant appearing in the denominator. As mentioned, the single infrared mass scale assumption could apply in medium in which case $f_\sigma^*\approx f_\pi^*$. This suggests that the $\sigma$ mass would stay more or less unscaled or even go up a little rather than down at some density.

\subsection{Mended symmetries}
 However, at high density very near the VM fixed-point, both the chiral-scale symmetry and the scale symmetry are supposed to be restored, $\beta'=0$ and $\la\bar{q}q\ra =0$ and the situation is changed dramatically such that  both the pion mass and dilaton mass are expected to vanish.
That the $\sigma$ tends, in the chiral-scale limit, to a massless excitation could be understood as reflecting its NG boson property manifested on the same footing as the pion. In the large $N_c$ limit, both are dominated by a $q\bar{q}$ component~\cite{CT}. In the same limit, the $\rho$ and $a_1$ tend to become massless joining the pion, revealing a generalized hidden local symmetry\cite{gen-hls}.  This multiplet structure of $\pi$, $\sigma$, $\rho$ and $a_1$ at the chiral transition could be interpreted as a case of  Weinberg's mended symmetries~\cite{weinberg}. As stressed in \cite{weinberg}, the non-derivative coupling associated with the scalar and the possible massless gauge fields $\rho$ and $a_1$ emerging at the critical point could play an important role in critical behavior of dense matter.
\subsection{How vector repulsion could win over scalar attraction?}
Let us consider what can happen as density exceeds that of nuclear matter by a large amount. Extrapolating the mean-field treatment given above for densities in the vicinity of nuclear matter, we look at the scalar and vector self-energies of the nucleon which can be given by the scalar and vector potentials, $S_N$ and $V_N$, respectively, gotten from the tadpoles of Fig.~1
\be
S_N&\approx& - \frac{{g_{\sigma NN}^*}^2}{{m_\sigma^*}^2} n_S\label{spotential}\\
V_N&\approx& \frac{{g_{\omega NN}^*}^2}{{m_\omega^*}^2} n_B\label{vpotential}
\ee
where $n_S$ is the scalar density $\propto \la\bar{N}N\ra^*$ and $n_B$ the baryon number density $\propto \la N^\dagger N\ra^*$. In applications to nuclei, $g^*_{\sigma NN} \sim 9-10$, $m^*_{\sigma}\sim 500-600$ MeV, $g^*_{\omega NN}\sim 13$ and $m^*_\omega\sim 700-800$ MeV. In medium, $n_S < n_B$, so one can see that there is a large cancellation between $S_N$ and $V_N$, leaving a small binding energy $B/m_N\sim 0.01$, essentially giving the QCD sum rule results.  A highly provocative consequence of this cancellation is that it can enable one to understand the tiny binding energy observed in medium nuclei, in a way similar to what the BPS skyrmion structure does as described in Section \ref{BPS}.

Assuming that the mean-field approximation holds at higher density than normal, we can make a simple estimate of what can happen as density increases. In compact-star matter, the mean field of the $\rho$ meson must enter, i.e, in the symmetry energy, but let us focus on the effects of the scalar $\sigma$ and the vector $\omega$. To be quantitatively accurate, the IDD (intrinsic density dependence) inherited from the matching needs to be accounted for as is seen in the vector-manifestation property of the $\rho$ meson but one can already see what's going on from what we have discussed above. There are two effects that enter here.
\begin{enumerate}
\item As density increases,  the scalar density gets increasingly suppressed relative to the baryon density.
\item The $\sigma$ mass could increase, albeit slowly, inversely proportionally to some power of $f_\sigma^*$ which could decrease, whereas the $\omega$ mass could drop proportionally to some power of $f_\sigma^*$. Thus, effectively, the scalar attraction will decrease while the vector repulsion will increase as density goes up above $n_0$.
\end{enumerate}
The consequence: {\it at some density, the repulsion will take over}. To pinpoint where this can happen would require more sophisticated calculation, taking into account the IDD in the parameters $\beta^\prime$, $v_2$ etc. ignored in the above estimation, but it seems highly plausible that it will happen. It will give rise to the changeover from a soft EOS to a stiff EoS at some density above $n_0$, typically at $\sim 2n_0$ as observed in the dense skyrmion matter.
\subsection{Solution to the ``hyperon problem"}
So far we have been considering nuclear systems without strangeness. The dilaton \`a la Crewther and Tunstall has a greater power when the strangeness flavor is considered, given that the mass scale of the dilaton is the same as that of the kaon. A particularly interesting question is the effective mass of hyperons in dense medium because it has to do with the EoS in compact-star matter. A standard approach to hyperons in star-matter is the relativistic mean field theory and this could be done with three-flavor HLS Lagrangian with baryons. In this approach, we can again resort to self-energies of the hyperons involved.

Whether or not hyperons enter in dense neutron-rich matter as in compact-star matter can be addressed in terms of the in-medium mass of hyperons. Most relevant to the issue is $\Lambda$, with the other hyperons less likely to figure. $\Lambda$s appearing at densities $\sim (2-3) n_0$ are understood to soften the EoS, with the consequence that $\sim 2$ solar-mass stars could not be stabilized. This is known as ``hyperon problem."

In our approach it suffices to consider, in the RMF approximation, the self-energy of the $\Lambda$ in neutron matter as in nuclear matter. It is given by the tadpoles of Fig.~1 with the external nucleon replaced by a $\Lambda$. The binding energy of a $\Lambda$ is estimated to be $\sim$ 28 MeV in nuclear matter~\cite{bedaque}, about 12 MeV bigger than that of a nucleon in nuclear matter. The only difference from the nucleon case is the couplings $g_{\omega \Lambda\Lambda}$ and $g_{\sigma \Lambda\Lambda}$ in place of $N$. We therefore expect that there will be the same changeover as in the nucleon case from attraction to repulsion at some density comparable to or perhaps somewhat higher than that of nuclear matter. It is also expected that the same mechanism would make $\Lambda$-$\Lambda$ interactions negligible. This could provide an extremely simple model for the mechanism put forward by Bedaque and Steiner -- taking place at $n\gsim 2n_0$ -- that could avoid the ``hyperon problem"~\cite{bedaque}. Numerically it is found to be in the range $1.5\lsim n/n_0\lsim 2.0$~ \cite{pklr}.
\section{Dense BPS matter}
\subsection{Skyrmion matter}
In the discussions made up to this point, baryons were explicitly included as matter fields. Alternatively -- and potentially more predictively, one can introduce baryons as skyrmions from  meson-only HLS Lagrangians. Particularly interesting is the Lagrangian obtained in holographic QCD obtained by Sakai and Sugimoto~\cite{sakai-sugimoto},  which is given by a 5D YM action in warped space plus a Chern-Simons action that encodes anomalies:
\be
S &=&  S_{YM} +S_{CS}, \\
S_{YM} &= &  \frac{N_c\lambda}{54\pi^3}\int \sqrt{-g}\frac{1}{8}  {\rm tr}
({\cal F}^2_{AB}) d^4x dz \label{YM} \\
S_{CS} &=& \frac{N_c}{16\pi^2}\int \hat{A}\wedge  {\rm tr} F^2 +\frac{N_c}{96\pi^2}\int \hat{A}\wedge {\hat{F}}^2.\label{CS}
\ee
Here $\hat{A}$ and $\hat{F}$ are the $U(1)$ component of $U(2)$ gauge field ${\cal A}$ and its field tensor, respectively, and unhatted quantities are $SU(2)$ components.  This action -- valid for large $N_c$ and large 't Hooft constant $\lambda=g_s^2N_c$ where $g_s$ is the QCD gauge constant -- has $U(2)$ local gauge invariance and supports {\it approximate} instanton that can be identified as the baryon. It works fairly well for baryon properties~\cite{HRYY}, particularly for vector-dominance structure of nucleon EM form factors that arises when dimensionally reduced \`a la Klein-Kaluza from 5D to 4D, with the resulting action consisting of pions and infinite towers of hidden local fields, i.e.,  iso-vector and iso-scalar vector mesons. Details, successes and failures, can be found in \cite{multifacet}.

In terms of large $N_c$ and large $\lambda$, the leading term is the ${\cal O} (N_c\lambda)$ term in the YM action (\ref{YM}) in flat space. The warping and the Chern-Simons term are subleading in $\lambda$, i.e., ${\cal O} (N_c \lambda^0)$. If one ignores the warping and the Chern-Simons term, then leading 5D YM action in flat space supports an exact instanton configuration with baryon quantum number 1. Reduced to 4D, the action will consist of the pion and the infinite tower of hidden gauge bosons, i.e., the iso-vector mesons, i.e., $\rho$, $\rho^\prime$, $\rho^{\prime\prime}$ ... and $a_1$, $a_1^\prime$, $a_1^{\prime\prime}$ ... The iso-scalar vector mesons, i.e., $\omega$ mesons, decouple and do not figure in the leading order. An extremely interesting observation made by Sutcliffe~\cite{sutcliffe} is that the skyrmion in the presence of only the iso-vector mesons approaches closer to  BPS skyrmion the more vector mesons are included in the Lagrangian. Applied in the limit of bringing in infinite vector mesons, the $A$-skyrmion system becomes a BPS state with zero binding energy for an $A$-nucleon nuclei. This gets closer to nature with very small binding energy for medium-mass nuclei. But this nice feature is obstructed by the presence of $\omega$ mesons. It was found that this approach to BPS is spoiled drastically once the ${\cal O}(\lambda^0)$ effects are taken into account~\cite{maetal13}. The warping of the space which comes in at that order already can seriously upset the BPS, but more importantly the lowest-lying $\omega$ meson alone can make the system deviate strongly from BPS as has been seen in numerical calculations of skyrmion matter on crystal lattice~\cite{maetal13}. It is not yet checked what happens when the infinite tower of $\omega$'s are introduced.
\subsection{Skyrmion crystal: Avoiding the $\omega$ disaster}
That something goes wrong in the presence of the $\omega$ meson has been noted when skyrmions are put on crystal lattice with the Lagrangian of the form
\be
{\cal L}_2 &\approx&
  {\calL}_{HLS}^{(2)} (\frac{\chi}{f_{0\sigma}})^2 + {\calL}_{hWZ}(\frac{\chi}{f_{0\sigma}})^{\delta}+\cdots \nonumber\\
   &+& \frac{f_\pi^2}{4} (\frac{\chi}{f_{0\sigma}})^3 {\Tr}(MU^\dagger +h.c.)\nonumber\\
    &+&\frac 12 \del_\mu\chi\del^\mu\chi + V(\chi)\label{cslagHLS}
 \ee
 where $V$ represents a generic dilaton potential. For illustration, only the quartic term in the hWZ Lagrangian is shown, others being subsumed in the ellipsis.  It is the hWZ terms that bring $\omega$ mesons coupling to other hadrons. In \cite{PRV1}, it was found that in the skyrmion matter constructed with (\ref{cslagHLS}) with $\delta=0$ had the properties such that the in-medium pion decay constant $f_\pi^*$ {\it increased} as density increased. Similarly the matter was stabilized only when the $\omega$ mass {\it increased}\footnote{This analysis was done with a Coleman-Weinberg-type dilaton potential, but the qualitative feature of the skyrmion crystal should not depend on the potential.}.

 Both of the above observations are totally at odds with nature. In particular the pion decay constant is expected almost model independently in theory (QCD) -- and experimentally observed up to normal nuclear matter density $n_0$ -- to decrease as density increases.

 However it was noted then that if one took, for no good reasons, $\delta\gsim 2$, both diseases were simply cured by an intricate interplay between the scalar $\sigma$ and the vector $\omega$~\cite{PRV}. Unfortunately the analysis of \cite{PRV} was found to be incomplete because the hWZ terms were truncated to only one term (out of three)  proportional to $\sim \omega_\mu B^\mu$ where $B_\mu$ is the baryon current -- and a full account of the three hWZ terms is found to be needed for quantitative accuracy~\cite{maetal13}. However, what is clear and noteworthy is that  a term of that type with $\delta=\beta^\prime$ is required to repair the disastrous results. It should be noted that this resolution would correspond to picking $c_{hWZ}=0$ and $\beta^\prime\gsim 2$ in (\ref{cslagM}). It is tempting to conjecture that the hWZ term reflecting the chiral anomaly carries  information on the explicit  scale symmetry breaking in medium encoded in the trace anomaly via the identification $\delta=\beta^\prime$.\footnote{Whether the magnitude $\beta^\prime\sim 2$ is reasonable or not is not clear. It is interesting however that in \cite{song}, a large positive anomalous dimension $\sim 2$ was found to be relevant, though in a different context. }
\subsection{BPS structure in heavy nuclei and dense matter?} \label{BPS}
The two observations, i.e., the emergence of the BPS structure with the infinite tower of vector mesons {\it  in the absence of the dilaton field and the $\omega$ field}, on the one hand, and the near cancellation of the strong scalar attraction and the strong vector repulsion leaving small binding energy in medium and heavy nuclei, on the other, must be closely related. We have interpreted these in terms of the interplay between the chiral symmetry and the conformal symmetry operative in medium. In some sense, related to these observations is, as mentioned, the suppression of the $\rho$ tensor forces in the symmetry energy so that the pion plays the dominant role in the EoS of asymmetric nuclear matter relevant to compact stars. {\it What results at high density therefore seems to be very simple: The dense matter is given by weakly interacting quasiparticles with only the pions intervening in the interactions.}

An extremely simple but puzzling case is the BPS matter for heavy nuclei constructed in \cite{adam}. There the main ingredient in the Lagrangian is a topological term quadratic in topological current, hence BPS, corrected by small terms such as Coulomb and iso-spin breaking. The model has no obvious connection to QCD. Yet, with a mild parameter adjustment, the model, with no role of  pions, is found to describe extremely well the Bethe-Weiz\"acker mass formula. This picture is a complete opposite to that of the Skyrme model which is built of  pions in a Lagrangian that encodes current algebras, thus accounting for low-energy nuclear interactions. However both have difficulties. While it has an indirect link to QCD as an effective field theory,  the skyrmion model with or without vector mesons fails badly to explain nuclear binding energy: The binding energy comes out much too big. On the other hand,  the BPS model, while surprisingly successful for nuclear binding energies,  with no obvious link to QCD, fail to describe pion-nuclear interactions. The question is: Can these defects in both models be repaired? This is a problem that begs to be resolved for the models anchored on topology to be viable. Nobody knows what the resolutions are. One possible conjecture for the BPS model is that one could have the pions couple to the BPS matter at the surface of the ``compacton," a bag containing the BPS matter, somewhat like the bag model for the nucleon, i.e., chiral bag model with the pions coupled at the surface with certain boundary conditions.
\section{Discussions}
In concluding this paper, we comment on a few issues that could be closely related to what we have discussed elsewhere or  remain poorly understood yet.

The structure of the dilaton and the $\chi$PT$_\sigma$ confirms the finding in the skyrmion-matter calculation of \cite{maetal13} that the bulk of the nucleon mass, $\sim 70\%$, comes from other than the vacuum realignment by density, i.e., partial restoration of  broken chiral-scale symmetry. It supports the thesis that the baryon mass is predominantly of gluon source. It does not explain, however, why the pion decay constant behaves also in the same way in density as the nucleon mass does, a relation that as observed in skyrmion matter, seems to hold in medium in the large $N_c$ limit. What this does is a surprising simplification of physics in high-density regime as mentioned above.

It follows almost trivially that $f^*_\sigma\approx f^*_\pi$ in dense matter~\cite{br91,maetal13}. We consider this as an indication that there is a single infrared mass scale for chiral symmetry and scale symmetry in medium.

Some of the issues that are not yet resolved or to be worked out are:
\begin{itemize}
\item It is unclear what the source of $\alpha_{IR}$ could be if it existed. So far it has not been detected in lattice QCD. And it may be difficult to pin-point the location of the fixed point. It cannot, however, be ruled out that such an IR fixed point could be present in highly correlated medium and is detectable. One possibility is that scale invariance emerges in a phase at certain density that is different from the phase at lower density that is continuously connected to the matter-free vacuum. An example is the half-skyrmion phase uncovered at high density, $\gsim 2n_0$, in dense skyrmion matter simulated on crystal lattice~\cite{half-skyrmion}. The question then is how to develop an expansion in $\beta^\prime$ in $\chi$PT$_\sigma$.
\item So far the treatment in dense medium was based on the assumption that the dense baryonic matter can be treated reliably by Fermi-liquid theory. This is the underlying reason why the density-functional approach as well as RMF theories are successful up to nuclear matter density. But the presence of the dilaton scalar, with non-derivative coupling, could induce baryonic matter to deviate drastically from Fermi liquid structure changing into non-Fermi liquid as density increases. Such non-Fermi liquid structure, { perhaps an ``un-Fermi-liquid" as in condensed matter physics~\cite{unparticle}}, could invalidate what has been taken for granted in the past.
\item Applied to nuclear matter, $\chi$PT$_\sigma$ could impact, via the scalar, on  many-body forces and hence on nuclear saturation, phase transitions  etc. Reformulating a systematic nuclear chiral-scale perturbation theory is an open problem.
\item RG studies showed that scalar exchange between nucleon and kaon is particularly important for kaon condensation. If the scalar involved is the dilaton associated with the IR fixed point structure, this feature will play a crucial role on kaon condensation~\cite{PR-kaoncon}. This is more so if the argument for the suppression of hyperons in compact-star matter discussed above is correct.
\end{itemize}

In this paper, we exploited the possible equivalence of the RMF approach to Landau Fermi-liquid theory and its power in addressing the notion of $f_0(500)$ as the scalar pseudo-NG boson $\sigma$ and in studying the interplay of the scalar $\sigma$ and the vector $\omega$ in high-density EoS. The arguments developed are quite general. In order to confront Nature, it is, however, more efficient to resort to the ``double-decimation approach" described in \cite{BR:DD} where the first decimation is made to obtain $V_{lowK}$ and then the second decimation is performed, using the $V_{lowK}$, to approach the Landau Fermi-liquid fixed point.  This method will be used in \cite{pklr} to have the general framework discussed in this paper confront Nature via the EoS of compact-star matter.

%%%%%%%%%%%%%%%%%%%%%%%%%%%%%%%%%%%%%%%%%%%%%%%%%%%%%%
\acknowledgments
We are grateful for valuable correspondences with Rod Crewther and Lewis Tunstall on their approach to the QCD dilaton and for discussions with Koichi Yamawaki on the technidilaton for Higgs. Part of this paper was written when two of the authors (HKL and MR) were visiting the Theory Group of RISP, IBS, for which we are grateful to Youngman Kim for support. The work of WGP is supported by the Rare Isotope Science Project of Institute for Basic Science funded by Ministry of Science, ICT and Future Planning and National Research Foundation of Korea (2013M7A1A1075764).

\end{document}